\documentclass[a4paper]{spie}  %>>> use this instead for A4 paper
\usepackage{a4wide}
\usepackage{amsmath,amsfonts,amssymb}
\usepackage{todonotes}
\usepackage{graphicx}
\usepackage{tabularx}
\usepackage{subcaption}
\usepackage{booktabs}
\usepackage{multirow}
\usepackage{enumitem}
\usepackage[colorlinks=true, allcolors=blue]{hyperref}
\usepackage{xcolor}
\usepackage{amsmath}
\usepackage{amssymb}
\title{Deep Learning-Based Carotid Artery Vessel Wall Segmentation in Black-Blood MRI Using Anatomical Priors}

\author[a]{Dieuwertje Alblas}
\author[a]{Christoph Brune}
\author[a]{Jelmer M. Wolterink}
\affil[a]{Department of Applied Mathematics, Technical Medical Centre, University of Twente, Enschede, The Netherlands}

\authorinfo{Further author information: (Send correspondence to D. Alblas)\\ D. Alblas: E-mail: d.alblas@utwente.nl}

\begin{document} 
\maketitle

\begin{abstract}

\noindent Carotid artery vessel wall thickness measurement is an essential step in the monitoring of patients with atherosclerosis. This requires accurate segmentation of the vessel wall, i.e., the region between an artery's lumen and outer wall, in black-blood magnetic resonance (MR) images. Commonly used convolutional neural networks (CNNs) for semantic segmentation are suboptimal for this task as their use does not guarantee a contiguous ring-shaped segmentation. Instead, in this work, we cast vessel wall segmentation as a multi-task regression problem in a polar coordinate system. For each carotid artery in each axial image slice, we aim to simultaneously find two non-intersecting nested contours that together delineate the vessel wall. CNNs applied to this problem enable an inductive bias that guarantees ring-shaped vessel walls. Moreover, we identify a problem-specific training data augmentation technique that substantially affects segmentation performance. We apply our method to segmentation of the internal and external carotid artery wall, and achieve top-ranking quantitative results in a public challenge, i.e., a median Dice similarity coefficient of 0.813 for the vessel wall and median Hausdorff distances of 0.552 mm and 0.776 mm for lumen and outer wall, respectively. Moreover, we show how the method improves over a conventional semantic segmentation approach. These results show that it is feasible to automatically obtain anatomically plausible segmentations of the carotid vessel wall with high accuracy. 
% In this work, we present a fully automatic two-step method for vessel wall segmentation of the , with representations of sub-voxel accuracy as a result. Furthermore, a  is enforced by an inductive bias in the method. The first step concerns local globalization, resulting in a continuous centerline of the carotid arteries. Ray casting around this centerline is performed in the consecutive step, yielding the exact radii of the lumen and outer wall at set angles, using a rotation-equivariant CNN. This results in a ring-shaped
% segmentation of the vessel wall.
% \textcolor{red}{
% \begin{itemize}
%     \item Vessel wall thickness important measurement for atherosclerosis
%     \item Current methods: either 'old school image processing', using active contour models, or CNN's producing voxel masks. Active contour methods rely on optimization of energy functions, which can experience difficulties in pathological regions or under low image quality. Vessel walls are very thin structures. Therefore sub-voxel precision preferred over voxel masks.
%     \item CNNs have no guarantee on the shape of the outcome of the prediction; our method incorporates inductive bias.
%     \item We present a two-step method with sub-voxel accuracy, using learned features for contour estimation.
%     \item Dice scores on cross-validation
%     \item Present a method for data augmentation, making the  contour estimation more robust.
% \end{itemize}}
%black blood MRI???
\end{abstract}


\section{INTRODUCTION}
\label{sec:intro}  % \label{} allows reference to this section
Atherosclerosis is one of the leading causes of morbidity and mortality worldwide. Characteristic pathology is a build-up of plaque in arteries such as the carotid artery that is visualized as a local thickening of the vessel walls, e.g., in black-blood magnetic resonance imaging (MRI). The thickness of the carotid artery wall is associated with the risk of incident stroke \cite{chambless2000carotid}, and thus, monitoring this thickness is an important aspect of the management of patients diagnosed with atherosclerosis. Performing this analysis manually is very time-consuming and prone to high inter- and intra-observer variability. Automating the segmentation of the carotid artery wall has the potential to improve atherosclerosis monitoring and management, and has been the subject of ongoing research~\cite{Wu2019DeepMRI,Chen2020AutomatedConversion}.

Automatic carotid vessel wall segmentation approaches typically involve two steps. First, the centerlines of the carotid arteries are located in a larger region-of-interest (ROI) encompassing the neck. 
% For this step, researchers can build on a large body of work on vessel tracking.
Second, in each axial slice, a local segmentation of the vessel wall is obtained in an ROI centered around the centerline. This second step involves fine-grained segmentation of a structure with a particular topology. Namely, the carotid artery wall is a ring consisting of two non-intersecting contours. This prior information has previously been exploited in segmentation methods using, e.g., ellipse fitting~\cite{adame2004automatic} or graph cuts \cite{arias2012carotid}. However, currently popular deep learning-based segmentation approaches using convolutional neural networks (CNNs) \cite{Wu2019DeepMRI} do not explicitly take this requirement into account and may lead to segmentation masks with holes or isolated voxels~\cite{Liu_2021}. 

In this work, we explore the idea of using polar coordinate systems for local ROIs and training CNNs to perform regression in these images~\cite{Wolterink2019GraphAngiography,10.1007/978-3-319-13972-2_13}. This means that we assume circular shape priors to describe the inner and outer contour of a vessel wall in terms of radii at equidistant angles around the centerline point. Here, similar to the work by Chen et al. \cite{Chen2020AutomatedConversion}, we nest two circles without intersection to obtain a ring-shaped segmentation mask. We use a rotation equivariant CNN that obviates the need for rotation augmentation, and thus uses the limited available training data more efficiently. Moreover, we experimentally compare single-slice (2D)  and multi-slice (3D) polar inputs. To account for possible irregularities in the initialization of the carotid centerline, we introduce a method-specific data augmentation technique that substantially improves the robustness of the method when little data is available. We evaluate our method in a public benchmark for carotid vessel wall segmentation and obtain the top-ranking result, both in terms of quantitative performance and qualitative performance~\cite{Challenge}.

\begin{figure}[t]
    \centering
    \includegraphics[width = \textwidth]{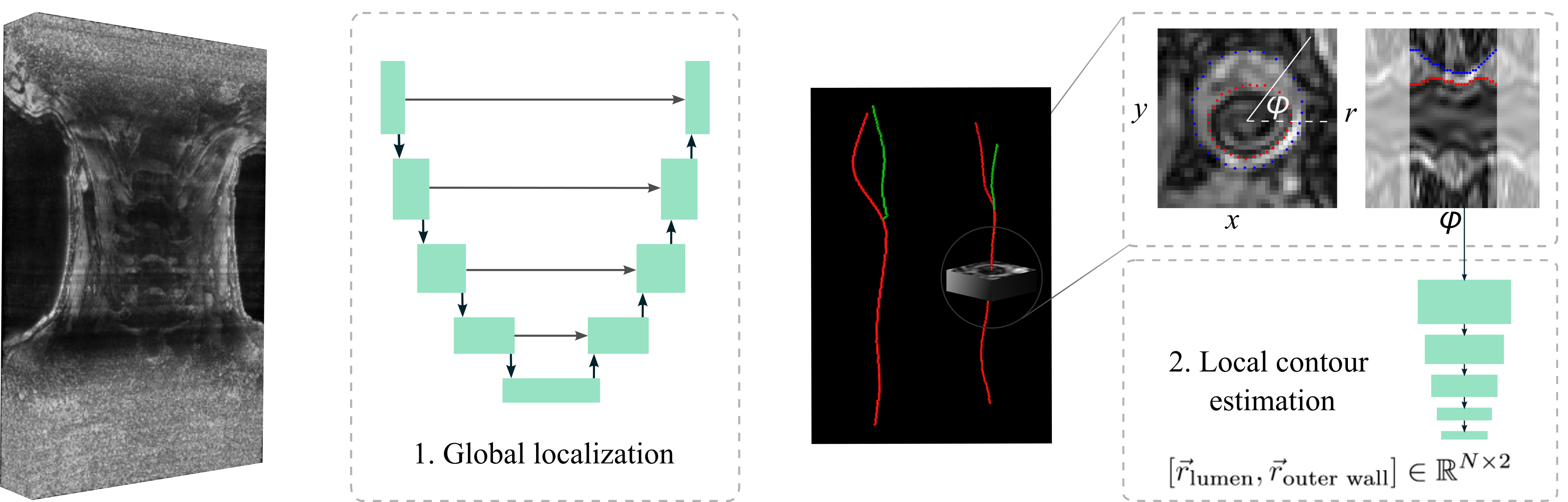}
    \caption{Overview of the proposed method. \textbf{Step 1}: centerlines of the common, internal, and external carotid arteries are extracted using Dijkstra's algorithm on a learned cost map $f$ (not pictured here). 
    %The centerline is extracted using Dijkstra's algorithm on $-f + \text{max}(f)$. 
    \textbf{Step 2:} local 3D regions-of-interest are extracted around the centerlines and transformed to a polar representation through ray-casting. A CNN predicts radii of the lumen and outer wall for each angle $\left[\phi_i \right]_{i=1,...,N}$.}
    \label{fig:pipeline}
\end{figure}

\section{DATA}
To develop our method, we used 26 black-blood MR images that are provided as a training set in the Carotid Artery Vessel Wall Segmentation Challenge~\cite{Challenge}. These images were acquired using a 3D Motion Sensitized Driven Equilibrium prepared Rapid Gradient Echo (3D-MERGE) sequence, as presented by Balu et al.~\cite{balu2011carotid}. The 3D-MERGE modality is optimized for visualization of high-risk, atherosclerotic lesions in carotid arteries and enables rapid scan-times at a high spatial resolution; acquisition requires two minutes, and yields isotropic, sub-millimeter voxel spacing. Our data set is a subset of the larger CARE II data set, acquired from 13 different hospitals and medical centers throughout China~\cite{zhao2017chinese}. This data set contains patients suffering from atherosclerosis in various degrees, resulting in a wide spectrum of vessel wall appearances. The MR image of each patient contains part of the common carotid artery, the carotid bifurcation, and part of the internal and external carotid artery, for the left and right side of the neck, respectively. Each MR image consists of 640 or 720 axial slices, with an isotropic voxel spacing ranging from 0.27 mm$^3$ to 0.39 mm$^3$.

For each axial image slice in this dataset, closed contours with sub-voxel accuracy of the lumen and outer wall are available in none or one of the following arteries: left internal and external carotid artery (ICAL/ECAL) and right internal and external carotid artery (ICAR/ECAR). In total, there are 2655 annotated slices, among which 2533 are ICAL or ICAR, and 122 are ECAL or ECAR, resulting in a major imbalance in the annotations between the internal and external carotid arteries. Since the axial slices were sparsely annotated, we manually annotated continuous lumen centerlines of the internal and external carotid arteries for all 26 patients, using the SimVascular software~\cite{updegrove2017simvascular}. We used these centerline annotations in the first step of our method.

% \textcolor{red}{
% \begin{itemize}
%     \item The data used to train both methods were 26 patients from the CareII data set \cite{zhao2017chinese}, as made available for the carotid artery vessel wall segmentation challenge.
%     \item 3D MRI black blood sequences, mostly 720 x 100 x 720 voxels, some exceptions of 640 x 160 x 640 voxels. Voxel spacing varied between 0.273 mm\textsuperscript{3} and 0.39 mm\textsuperscript{3}. Resampled to 0.5 mm\textsuperscript{3} for training the network used in the first step.
%     \item Outer wall and lumen contour annotations on non-consecutive slices. 
%     \item For the global localization step, centerlines of carotid arteries were annotated manually in SimVascular.
%     \item Splits for 5-fold cross validation, in order to get realistic performance measurements.
% \end{itemize}}

% Contour points of the lumen and vessel wall were given on a non-continuous sequence of axial slices in the scans. \textcolor{red}{Data imbalance?} In order to obtain realistic performance measurements, k-fold cross validation has been used, dividing the data set into five different train/test splits, each of 21 and 5 train and test patients respectively. Additionally, for training the network used in the global localisation step of the segmentation, annotations were made using SimVascular \textcolor{red}{ref???}.

\section{METHOD}\label{sec:method}
We follow the commonly used two-step approach for segmentation of the carotid vessel wall described in the Introduction, and shown in Figure \ref{fig:pipeline}. First, centerlines of the left and right internal and external carotid arteries are \textit{globally} located in the 3D MRI image. Second, the lumen and outer vessel wall contours are \textit{locally} estimated.

\subsection{Centerline localization}\label{sec:global}

To locate the centerlines in the first step, we define a cost-function $f(I): \left[0, 1 \right]^{m \times n \times p} \rightarrow \mathbb{R}_+^{2 \times m \times n \times p}$ on the MRI image~\cite{Sironi2016MultiscaleDetection}. This function describes the proximity of each voxel to the centerlines of the external carotid artery, and the internal carotid artery, which includes the common carotid artery. $f(I)$ is nonzero within a predefined radius of the centerline and decreases exponentially as the distance to the centerline increases: 

\begin{align}\label{Sironi}
    f(I_x) = \begin{cases} e^{a\left(1-\frac{D_C(x)}{d_M}\right)} - 1 \hspace{1cm} &\text{if } D_C(x) < d_M\\
    0 &\text{otherwise},
    \end{cases}
\end{align}
where $f(I_x)$ is the value of $f(I)$ at location $x$, $D_C$ is the metric distance between $x$ and the centerline, $d_m$ is a predefined nonzero radius, and $a$ is a control parameter, that we set to 6, as was recommended by Sironi et al.~\cite{Sironi2016MultiscaleDetection}.

The two channels of $f(I)$ represent the proximity maps for the centerlines of internal and external carotid arteries separately. A 3D U-Net is trained to predict $f(I)$, based on the MRI data\cite{cciccek20163d}. We use the symmetry in the sagittal plane to simplify the learning problem; right halves of the images are flipped and used for training, effectively reducing this problem into centerline localisation for the left carotid arteries. At inference, we likewise flip the image to predict $f(I)$ on the right side of the patient, and consecutively flip the network's prediction to obtain bilateral centerline proximity maps.

Finally, the centerline of each carotid artery is approximated by finding the shortest path between the first and last image slice using Dijkstra's algorithm with a cost-function based on the centerline proximity maps. By Equation \eqref{Sironi}, $f(I)$ attains its maximum on the centerline. However, for successful tracing of the centerline using Dijkstra's algorithm, a local minimum is required here. Moreover, our implementation of Dijkstra requires positive values of $f(I)$. Hence, the cost-function the algorithm operates on is: $-f(I) + \text{max}(f(I))$. We use the peak in the proximity map each 50 slices as start- and endpoints to increase robustness. The main advantage of our approach is that a continuous centerline is guaranteed to be found, in contrast to, e.g., a previously proposed 2D tracking-based method for the carotid artery centerline that connects independently determined 2D points of interest\cite{Chen2020AutomatedConversion}. %Since they connect independently determined 2D points of interest, disconnected sections of the centerline can emerge. 
% Due to extensive use of axial context information, our network learns the differences between the internal and external carotid arteries, producing centerlines for both.

\subsection{Wall segmentation}
In the second step of our method we locally estimate the vessel wall.
In each axial image slice, we estimate two closed nested contours that correspond to the lumen and outer artery wall. We describe both contours in polar coordinates in terms of local radii at fixed angles $\left[\phi_i\right]_{i=1,...,N}$ and use a multi-task CNN to simultaneously estimate radii values for both contours at each angle based on image data. To obtain non-intersecting contours, we define $r^i_{\text{outer wall}}$ as the sum of $r^i_{\text{lumen}}$ and the local vessel wall thickness, and let the CNN predict the latter, so that $r^i_{\text{lumen}} \leq r^i_{\text{outer wall}}, \forall i \in [0,N]$. The input to the CNN is a polar image, constructed by taking the centerline point intersecting an axial slice as the origin. This polar image can be easily constructed from the original Cartesian data using ray-casting at $N$ angles. At each angle $\phi_i$, a ray with a fixed length is cast, resulting in a vector containing the voxel intensities the particular ray passes in the Cartesian image. Rays are stacked to form a 2D polar image. Subsequently, this image is used as input to a CNN that predicts the aforementioned radii values $r^i_{\text{outer wall}}$ and $r^i_{\text{lumen}}$. In contrast to widely used CNNs for semantic segmentation, this CNN does not contain pooling layers and is thus translationally equivariant. Instead, to rapidly aggregate sufficient context information from neighboring angles and along image rays, it uses dilated convolution kernels. As in \cite{yu2015multi}, the dilation rate in this network doubles in each layer, allowing an exponentially growing receptive field with a linearly increasing number of trainable parameters. Because the network is translationally equivariant in the polar representation, it is rotationally equivariant in the Cartesian representation. In other words, rotating the input image leads to an equivalently rotated contour. Because rotation is periodic, we concatenate the first and last half of rays to the back and front of the polar image, respectively. This is visualized by the transparent boxes in Figure \ref{fig:pipeline} and results in a final polar image $I_{polar} \in \left[ 0, 1 \right]^{(2N-1) \times R}$. Exploiting the rotational equivariance of our network's architecture and the periodicity of the polar images obviates the need for rotational augmentation during training or testing as is done in Chen et al.~\cite{Chen2020AutomatedConversion}.

\begin{figure}[t]
    \centering
    \includegraphics[width=\textwidth]{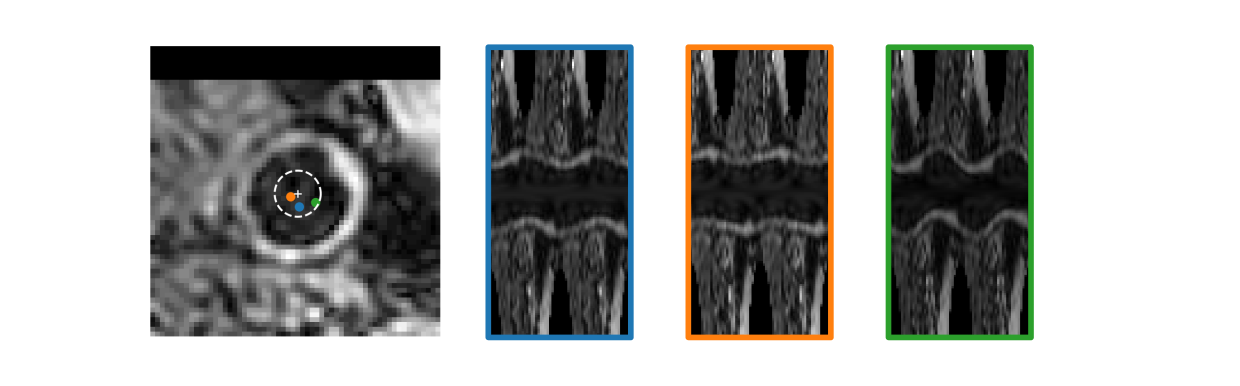}
    \caption{Proposed method for problem-specific data augmentation. Center points are randomly sampled on an axial slice (left), within a predefined region around the ground-truth center of mass of the lumen, indicated by the white `+'. The resulting, slightly different, polar images are shown on the right.}
    \label{fig:aug_procedure}
\end{figure}

The appearance of the carotid artery wall in adjacent image slices is typically correlated, and thus, adjacent slices may provide useful context information for segmentation of the artery wall. We evaluate both a single-slice and a multi-slice version of this CNN to obtain vessel wall contours in a single axial slice. In the single-slice version, 2D polar images are acquired as previously described. In the multi-slice version, 2D polar images are stacked with polar images constructed from  adjacent slices in both directions to obtain a 3D polar image. The architecture of the dilated CNN is adapted to take this additional context information into account.
    
Polar images are extracted based on a centerline point in the lumen. For slices with annotated contours that we use as training samples, this point is perfectly centered in the lumen. However, for new and unseen data, limitations in the centerline localization step may lead to incorrectly centered polar images. To compensate for this and improve robustness, we simulate this phenomenon in the training data by randomly sampling the centerpoint of the polar image within a small radius of the center of mass. This results in a number of slightly different polar images, which are used to train the CNN, as shown in Figure \ref{fig:aug_procedure}.

To demonstrate the effectiveness of the second step of our method, we compare our CNN to a 2D U-Net operating on Cartesian input images. The input images for this network are image patches, centered on the centerline point on the axial slice as determined by our global localization step. This network has a two-channel output with pixel-based segmentations: the first channel for the area within the vessel lumen, the second one for the area within the outer wall. We acquire the ring-shaped vessel wall by subtracting the lumen from the outer wall prediction.

\section{EXPERIMENTS AND RESULTS}
% We resampled all images to an isotropic voxel spacing of 0.5 mm$^3$. 

\subsection{Experimental settings}
Our data set consisting of 26 patients from the training set of the Carotid Artery Vessel Wall Segmentation Challenge~\cite{Challenge} was separated into five folds for cross-validation. As a general pre-processing step, all voxel intensities were rescaled to a $[0,1]$ range between the 5\textsuperscript{th} and 95\textsuperscript{th} intensity percentiles of each individual image. The following steps were repeated for each of the five folds.

In the first step of our method, a 3D U-Net was trained to predict the cost-image $f(I)$ for centerline localization. To this end, all images were resampled to an isotropic voxel spacing of 0.5 mm$^3$. We used randomly sampled patches of $368\times64\times320$ voxels mini-batches of eight samples to train the U-Net. The U-Net was optimized using an Adam optimizer for 400 epochs, with a learning rate of 0.001, multiplied by 0.1 in epochs 100 and 300 in combination with an MSE loss function. During inference, we processed the entire MRI image at once using the trained network as a sliding window. Dijkstra's algorithm was applied as described in Section \ref{sec:global} and coordinates of the resulting centerlines were expressed in world coordinates. These centerlines served as input to the second step.

\begin{figure}[t!]
\centering
    \includegraphics[width = \textwidth]{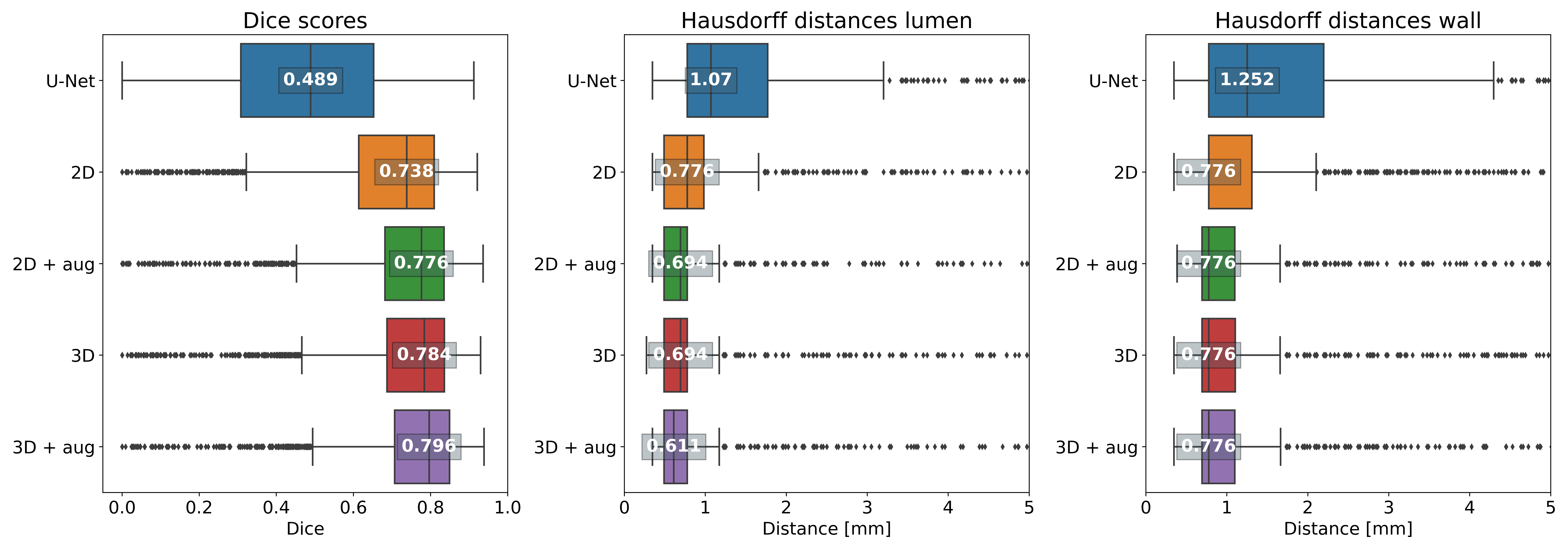}
        \caption{Performance metrics (Dice similarity coefficient and Hausdorff distances for lumen and outer wall) for U-Net baseline, single-slice and multi-slice polar images $I_{polar}$, with and without training data augmentation.}
    \label{fig:boxplot}
\end{figure}

In the second step of our method, contours of the lumen and outer wall were locally estimated. This step was performed on our original images, i.e., no global image rescaling as in step one was performed. Here, we compare single-slice and multi-slice CNNs, both with and without data augmentation, for a total of four models. In all cases, we used $N=31$ equidistant angles $\phi_i$ to parametrize contours and extract polar images. Rays in these polar images had a length of 127 voxels and a spacing varying between 0.21 mm and 0.31 mm. For CNNs with multi-slice inputs, 3 adjacent slices were included in both directions. We trained the CNNs on mini-batches of 100 polar images, with an Adam optimizer for 200 epochs, a learning rate of 0.001 and a smooth L1 loss function.

In addition to the aforementioned variants of our method, we trained a 2D U-Net for local contour estimation using semantic segmentation as a baseline model. The U-Net operates on $64\times 64$ pixel Cartesian image patches centered around the centerline acquired in the first step. Hence, for each of the polar images used by our single-slice model, there is an equivalent Cartesian image used by the U-Net model. %The U-Net model outputs two channels: one voxel mask for the lumen, and one voxel mask for the outer wall, including the lumen. The vessel wall is found by subtracting the lumen from the outer wall voxel mask. 
We trained this network for 300 epochs using a Tversky loss function, with $\alpha = 1$ and $\beta = 2$. For optimization of the network, we used an Adam optimizer with a learning rate of 0.001. During training, we used random rotations and random croppings of $64\times64$ within a predefined region around the ground truth center of mass of the lumen as data augmentation. 

\begin{figure}
\begin{tabular}{c c c}
    U-Net baseline & 2D polar + aug & Ground truth \\ 
    \includegraphics[width=0.32\textwidth, trim = {2cm, 3cm, 2cm, 2cm}, clip]{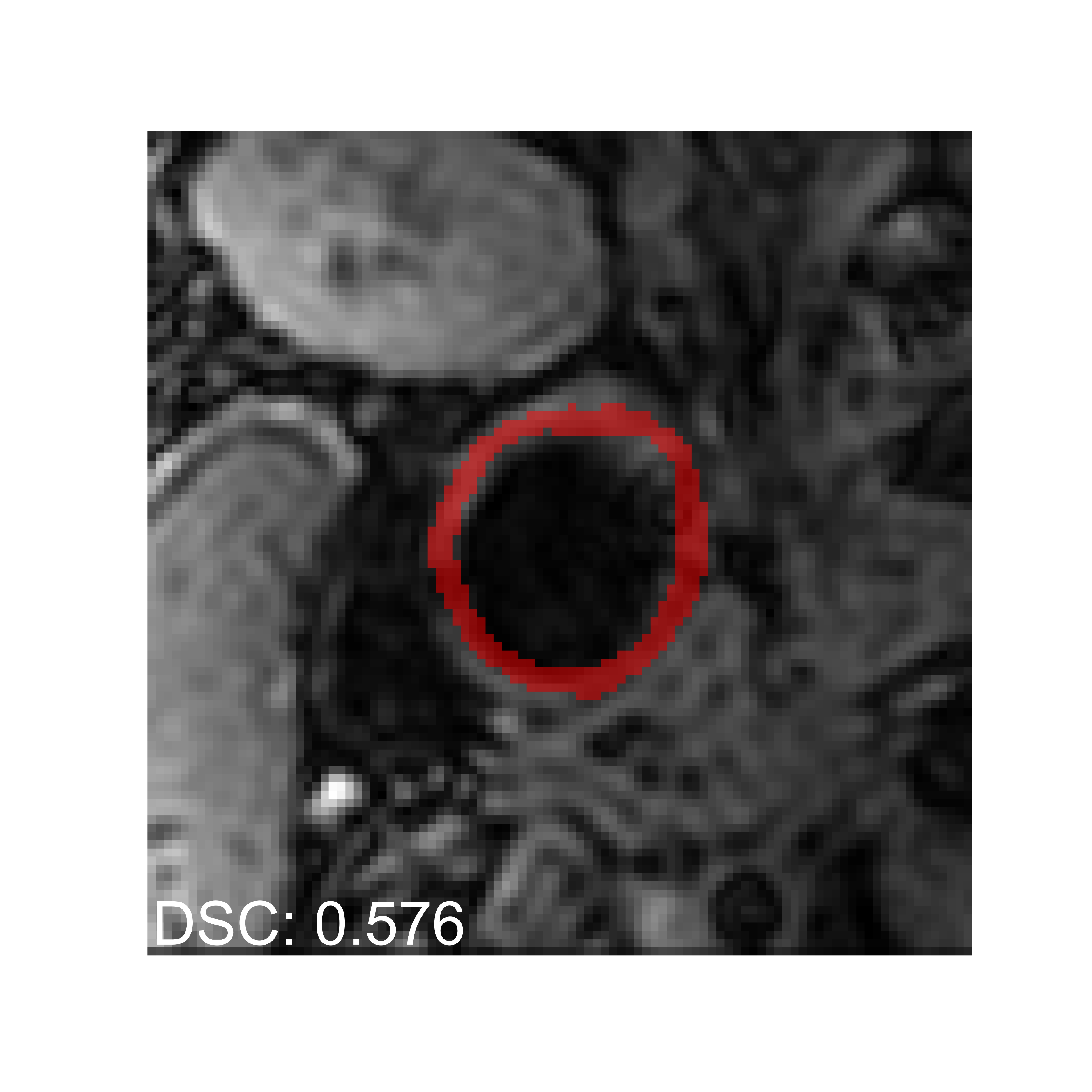} & 
        \includegraphics[width=0.32\textwidth, trim = {2cm, 3cm, 2cm, 2cm}, clip]{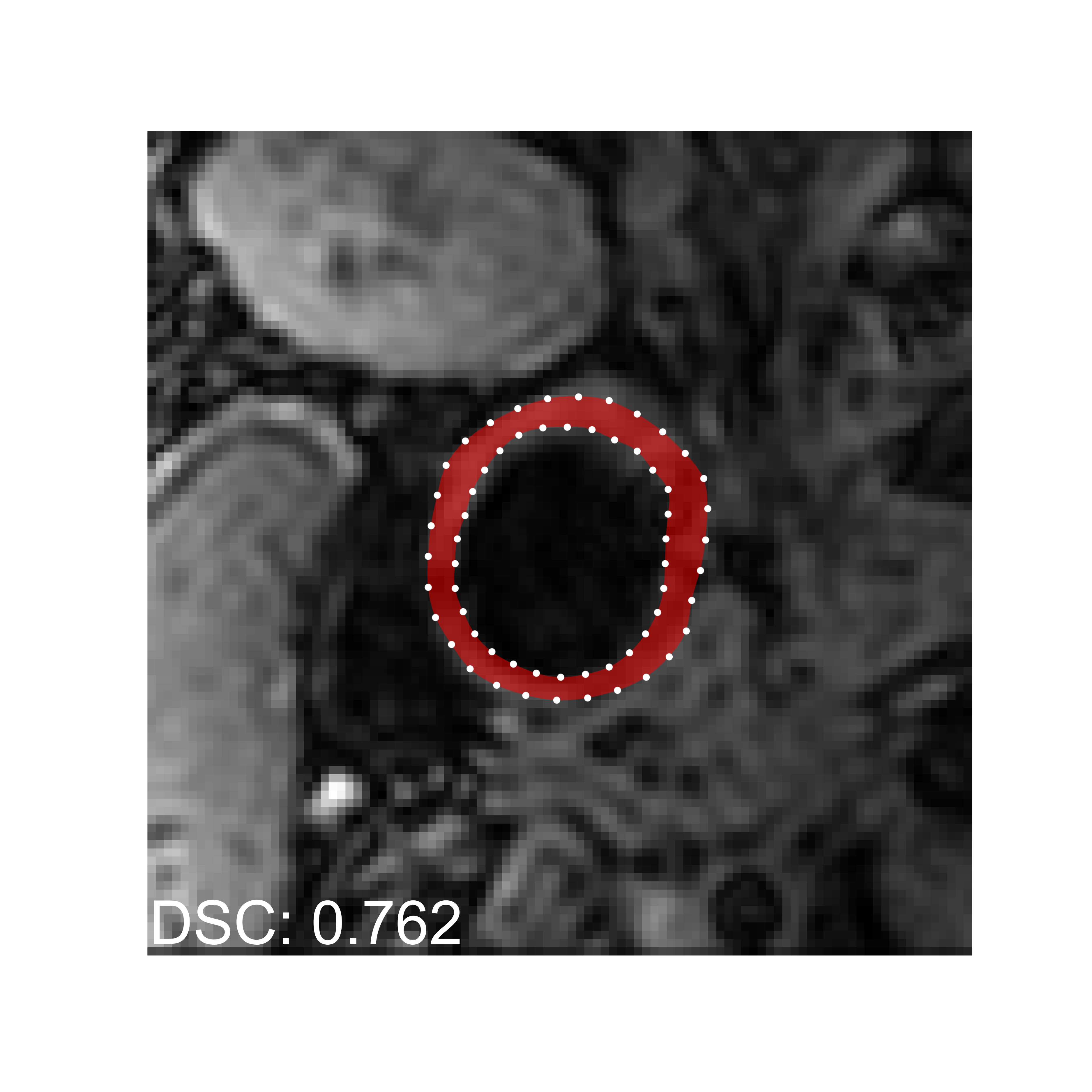}
       & \includegraphics[width=0.32\textwidth, trim = {2cm, 3cm, 2cm, 2cm}, clip]{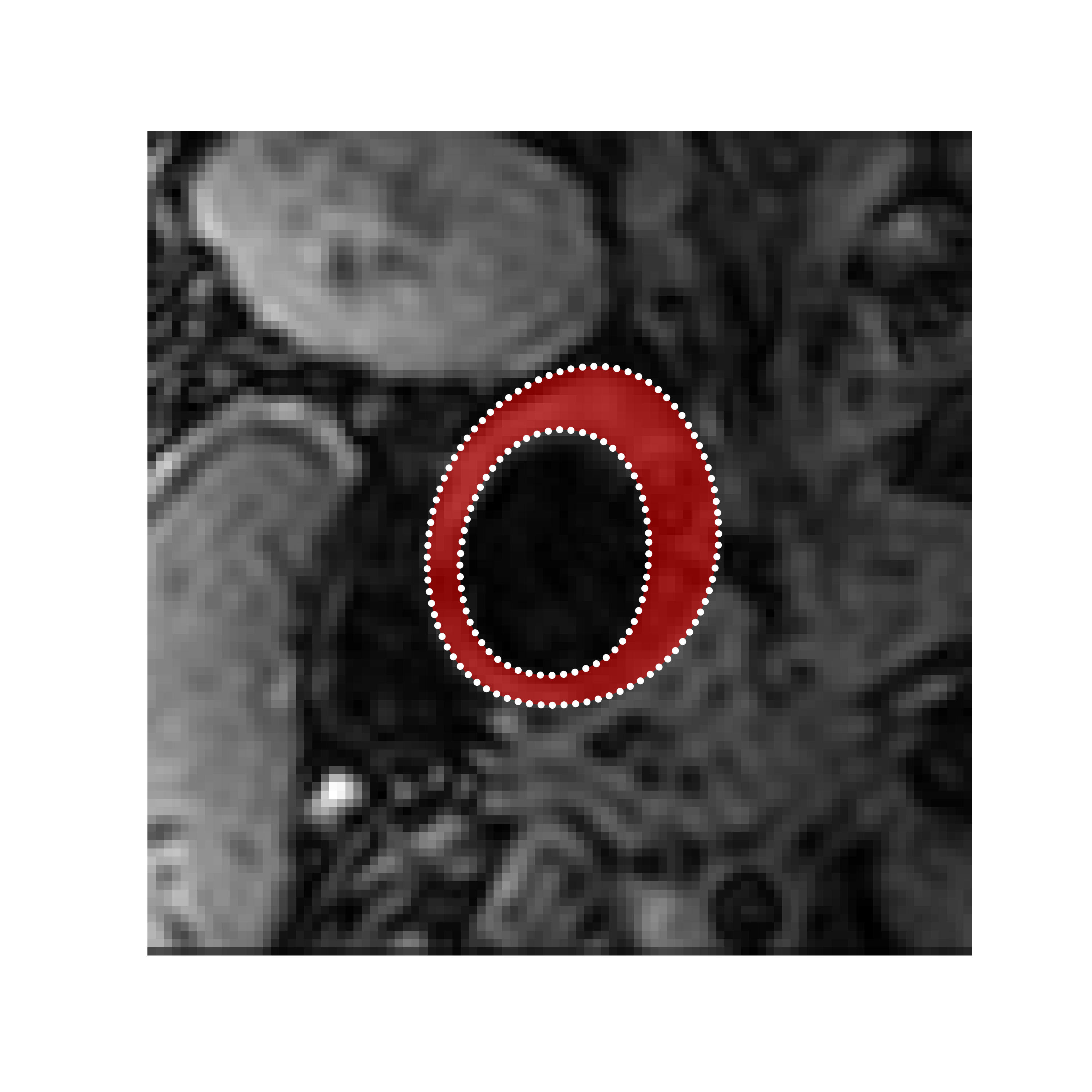} \\ 
    \includegraphics[width=0.32\textwidth, trim = {2cm, 3cm, 2cm, 2cm}, clip]{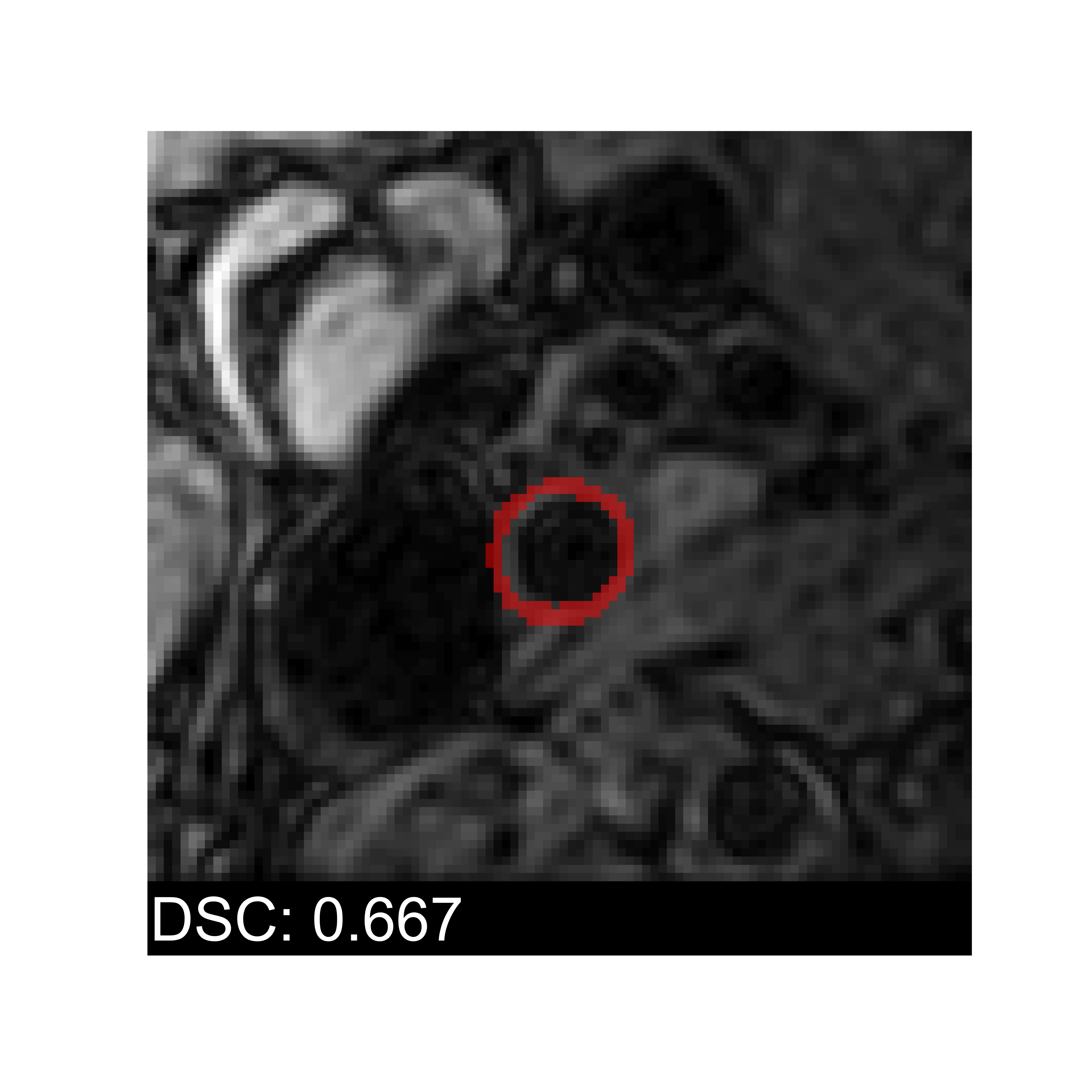}
        & \includegraphics[width=0.32\textwidth, trim = {2cm, 3cm, 2cm, 2cm}, clip]{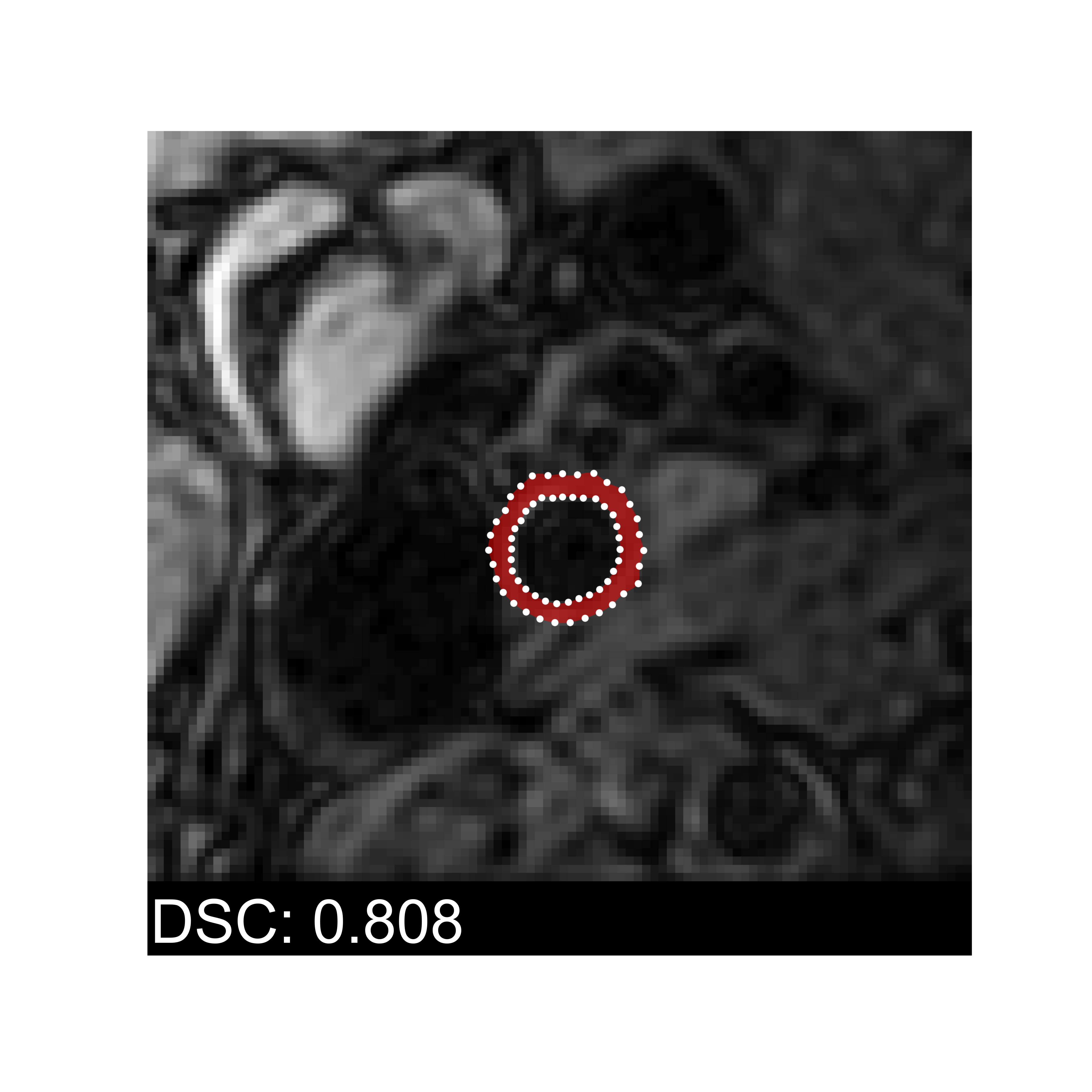}
         & \includegraphics[width=0.32\textwidth, trim = {2cm, 3cm, 2cm, 2cm}, clip]{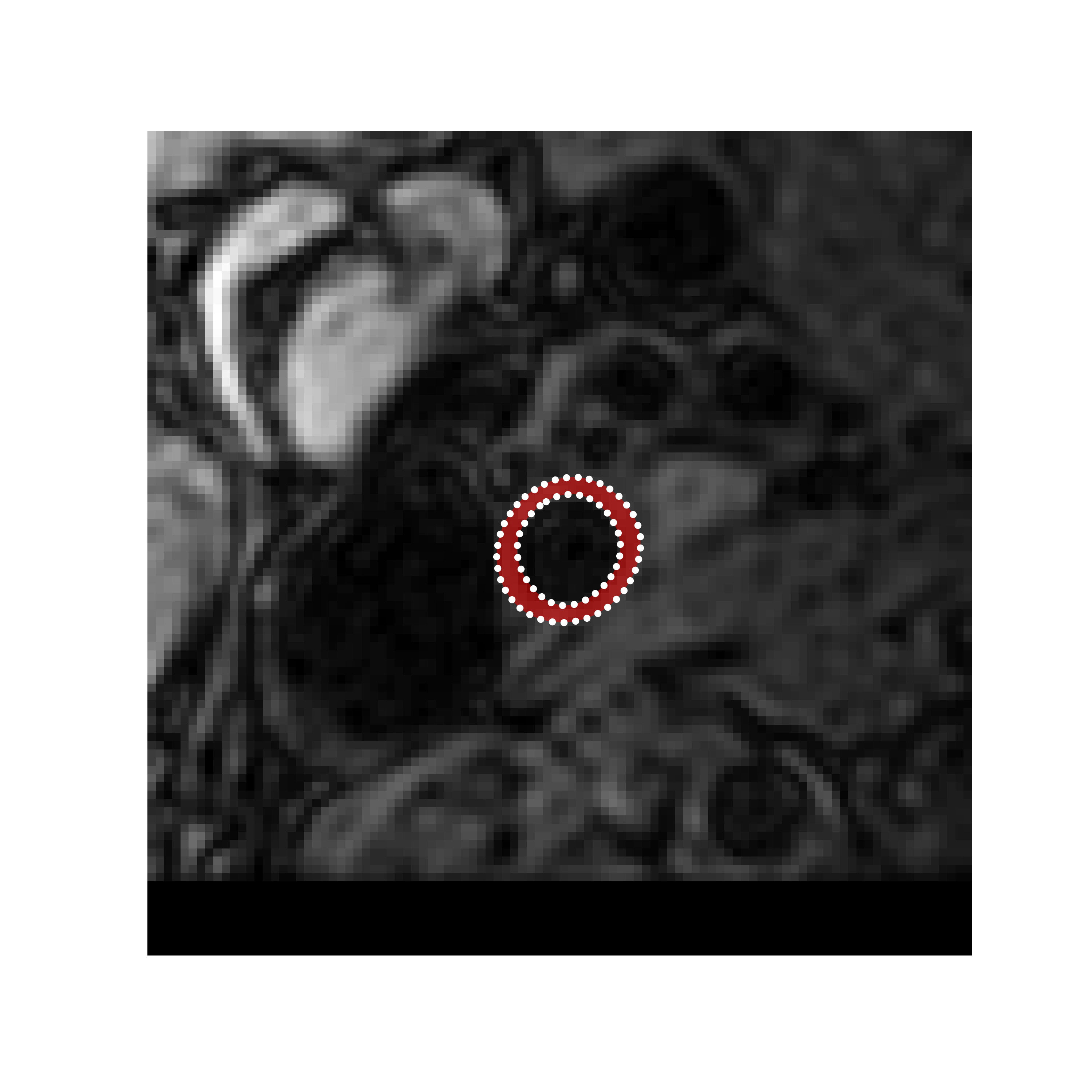}\\
    % \end{subfigure}
    \end{tabular}
    \caption{Vessel wall segmentations for variations of the second step in our automatic method: U-Net baseline (left), polar 2D images with augmentation (middle) and ground truth (right).}
    \label{fig:comparison_Unet}
\end{figure}

\subsection{Quantitative results}
We assessed the segmentation performance of our method in terms of the Dice similarity coefficient (DSC) and Hausdorff distance (HD) on both contours, following the criteria used in the Carotid Artery Vessel Wall Segmentation Challenge~\cite{Challenge} and using the evaluation scripts provided by the challenge organizers.

Figure \ref{fig:boxplot} shows quantitative results for our five-fold cross-validation experiments comparing the five evaluated models. First, the results show that operating on polar images with a rotationally equivariant CNN is a very effective method for vessel wall segmentation. The network trained on 2D polar slices without augmentation shows a considerable improvement compared to the 2D U-Net approach using the same image information. This difference is visualized in  Figure \ref{fig:comparison_Unet}. It becomes clear that the U-Net has some irregularities around the edges of the vessel wall, and does not perform well on vessels with a lot of plaque build-up. While the 2D polar approach also leaves room for improvement in this example, it shows a better performance at distinguishing the vessel's lumen.

%where segmentation results at the carotid bifurcation are shown. This is a challenging area to segment since the two nested shapes are non-convex. For the polar approach we managed to overcome this by extending the centerlines of both the internal and external carotid arteries into the common carotid arteries and merge the circles if the lumens intersect. For the U-net approach, this is considerably harder to deal with, as is seen in Figure \ref{bif_Unet}. Adapting this method likely requires a lot more data around the bifurcation in order to provide accurate segmentations there. Moreover, in slices where a lot of plaque has accumulated on the vessel wall, such as in Figures \ref{tvw_Unet} and \ref{tvw_3D}, the U-net shows poor performance compared to our proposed method, and the median HD is 0.42. 

\begin{figure}[t]
    \centering
        \begin{subfigure}{0.2\textwidth}
    \centering
    \includegraphics[height = 3.5cm, trim = {1cm, 0, 3cm, 0}, clip ]{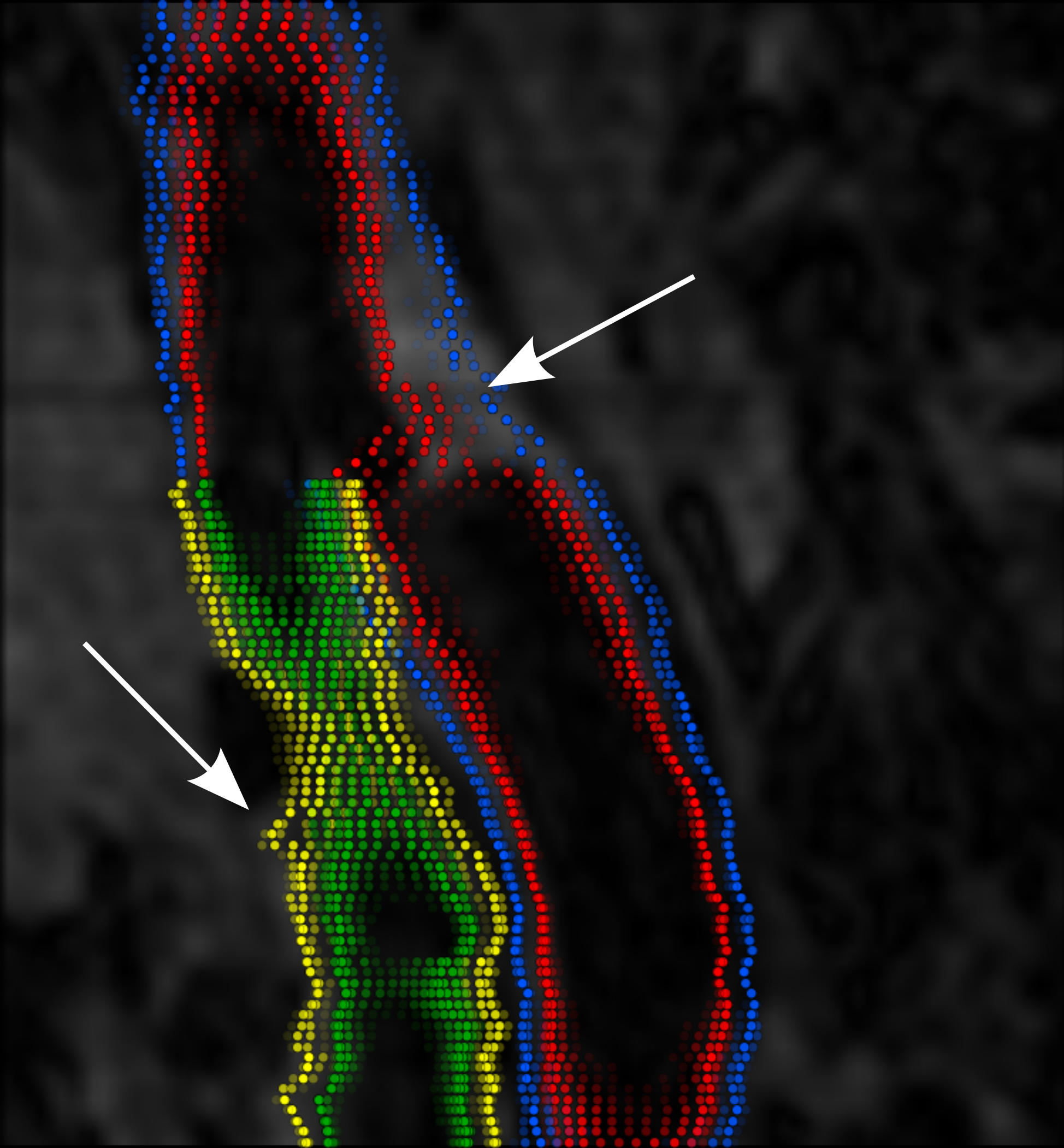}
    \caption{2D + aug} 
    \label{coronal_2D}
    \end{subfigure}\begin{subfigure}{0.2\textwidth}
    \centering
    \includegraphics[height = 3.5cm, trim = {1cm, 0, 3cm, 0}, clip]{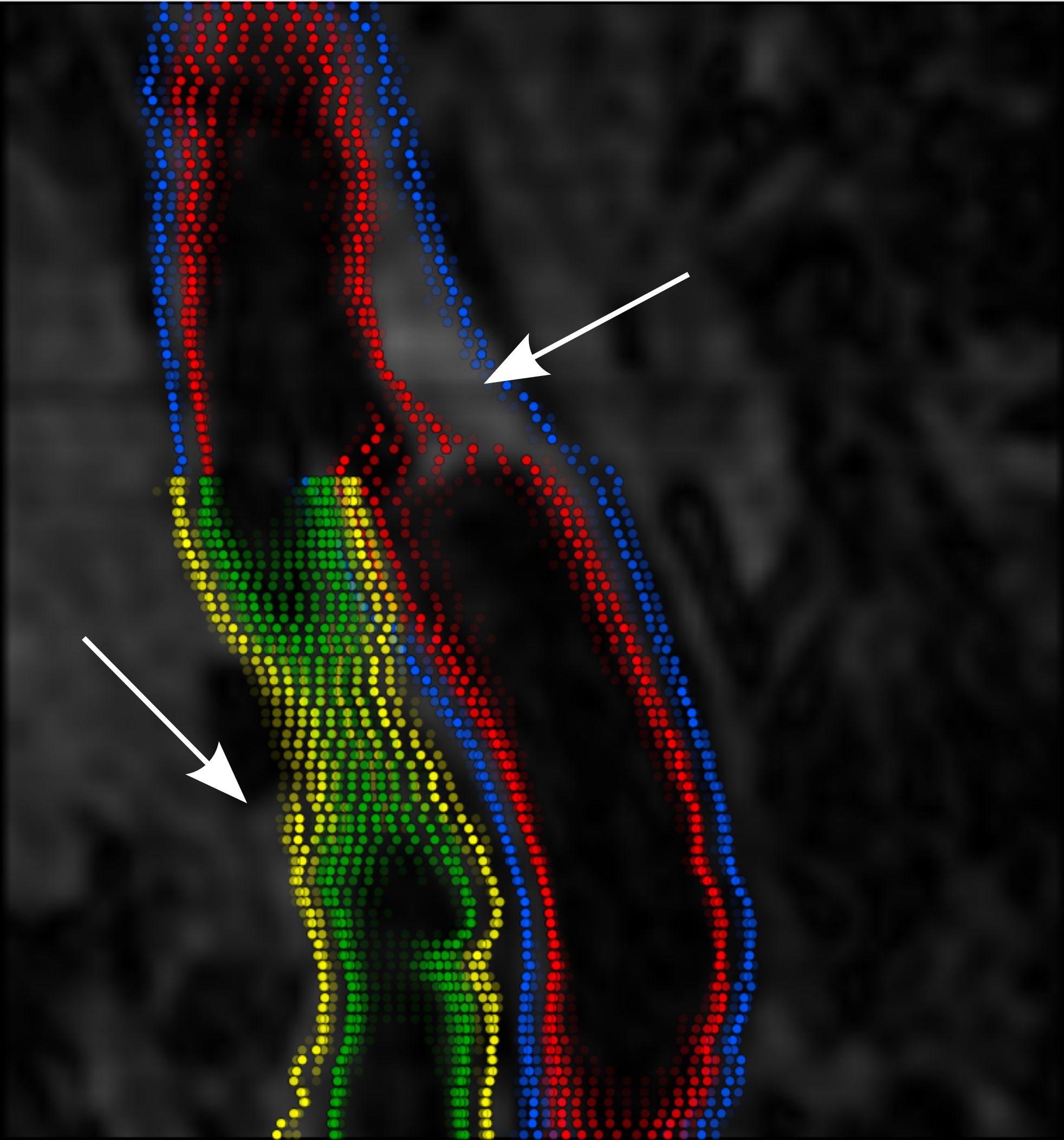}
    \caption{3D + aug}
    \label{coronal_3D}
    \end{subfigure}\begin{subfigure}{0.25\textwidth}
    \centering
        \includegraphics[height = 3.5cm]{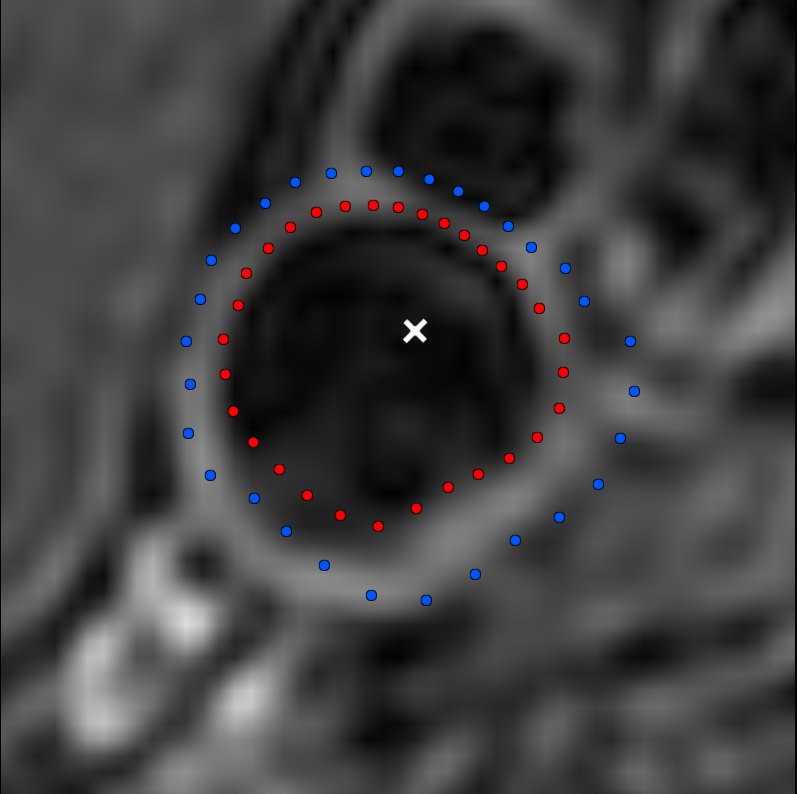}
        \caption{2D}
        \label{axial_noaug}
    \end{subfigure}\begin{subfigure}{0.25\textwidth}
    \centering
    \includegraphics[height= 3.5cm]{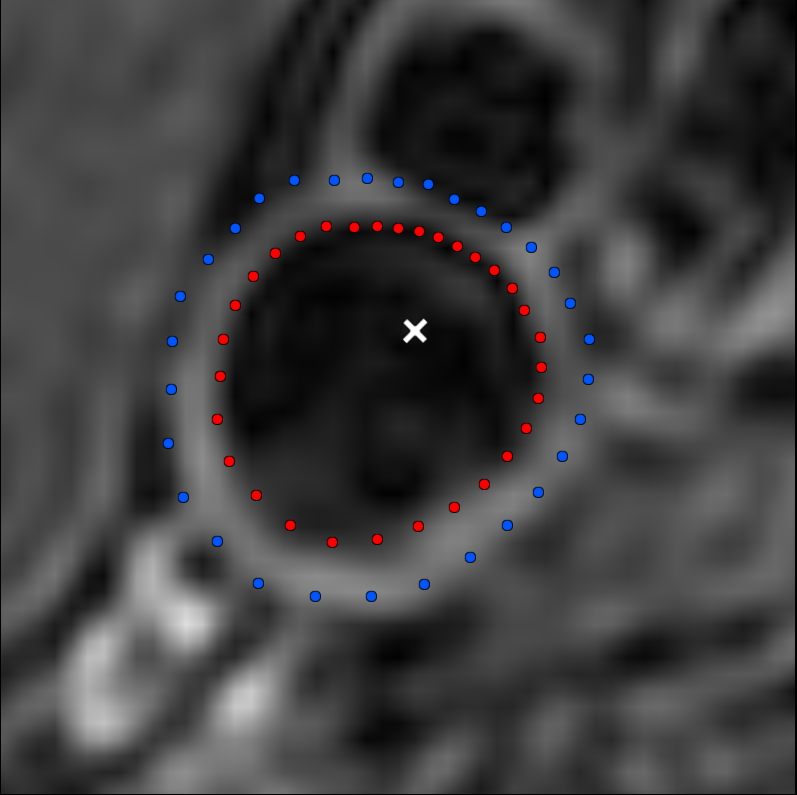}
    \caption{2D + aug}
    \label{axial_aug}
    \end{subfigure}
    \caption{MRI-images with automatically detected contours for ICAR (red and blue) and ECAR (green and yellow).
    (\subref{coronal_2D}) and (\subref{coronal_3D}) show a coronal cross-section near the bifurcation. (\subref{axial_noaug}) and (\subref{axial_aug}) show the effect of training data augmentation in an axial slice for which the centerline point is off-center.}
    \label{fig:segs_example}
\end{figure}

Additionally, Figure \ref{fig:boxplot} shows that networks operating on multi-slice or 3D polar images offer an improvement over the single-slice approach in terms of DSC values. This is likely due to the additional context information provided in adjacent slices. Figures \ref{coronal_2D} and \ref{coronal_3D} show the effect of using 3D polar images instead of 2D in the bifurcation area visualized in a coronal view. The red and blue contours show the lumen and outer wall contours of the right internal carotid artery, and likewise, green and yellow for the right external carotid artery. It becomes clear that the use of context in the axial direction results in an improved consistency of the contours, especially in the areas with a lot of plaque.

Finally, results show that the proposed problem-specific data augmentation is beneficial for both 2D and 3D polar images, but has the most impact on the 2D case, resulting in an increased median DSC from 0.738 to 0.776, and an improved HD from 0.776 to 0.694 mm in the lumen. Figures \ref{axial_noaug} and \ref{axial_aug} display the effect of augmentation for 2D polar images. The automatically identified centerline, marked by the white cross, is far off-center. The CNN trained on augmented images shows segmentations that match the underlying image better than the non-augmented one.

% \begin{table}[h!]
% \caption{Median Dice scores on segmentation of the vessel wall of our method on validation sets of 5-fold cross validation, with and without use of data augmentation in ray-casting.}

%     \begin{tabular}{l p{2cm} p{2cm} p{2cm}}
%     \toprule
%   & \textbf{Dice score} & \textbf{Hausdorff lumen [mm]} & \textbf{Hausdorff outer wall [mm]} \\
%     %  \cmidrule(lr){1-1} \cmidrule(lr){2-2}  \cmidrule(lr){3-3} \cmidrule(lr){4-4} \cmidrule(lr){5-5}
%     \cmidrule(lr){1-4}
%   2D & $0.738 \pm 0.196$ & $0.776\pm0.491$  & $0.776\pm 0.533$ \\
%   2D + aug & $0.776 \pm 0.153$ & $0.694 \pm 0.285$ & $0.776 \pm 0.153$\\
%   \cmidrule(lr){1-4}
%   3D & $0.784 \pm 0.149$  & $0.694 \pm 0.285$  & $0.776 \pm 0.410$   \\
%   3D + aug & $0.796 \pm 0.142$ & $ 0.611 \pm 0.285$ &  $ 0.776 \pm 0.403$\\ 
%     \bottomrule
%     \end{tabular}
%     \label{tab:results_aug}
% \end{table}

%\section{NEW OR BREAK-THROUGH WORK TO BE PRESENTED}
%This work shows how anatomically plausible and accurate vessel wall segmentation can be obtained in black-blood MRI images using a polar image representation and problem-specific data augmentation.

\section{CONCLUSION AND DISCUSSION}
We have presented a method that accurately segments the carotid artery vessel wall in black-blood MRI images. The method uses a polar conversion of local image patches to predict contours with a translationally equivariant CNN, leading to rotational equivariance in the original image space and a more effective use of training data. We demonstrated that our approach outperforms a semantic segmentation method that operates on Cartesian images. Moreover, we introduced the use of multi-slice polar images and problem-specific augmentation to improve the performance of our method. Both adjustments resulted in an improved segmentation performance over the non-augmented single-slice approach.

Here, we have presented a systematic comparison of experimental conditions on the \textit{training} set of the Carotid Artery Vessel Wall Segmentation challenge. In addition, we evaluated our method on the \textit{test set} of the public challenge, using an ensemble of the five networks trained for cross-validation. For this evaluation, we used our best performing model, namely the multi-slice model with augmentation. Quantitative results indicate an improved segmentation performance compared to the single networks reported in Figure \ref{fig:boxplot} and ranked highest in the competition (median DSC 0.813, HD lumen 0.552 mm and HD wall 0.776 mm). %With this , we demonstrated that it is feasible to obtain anatomically plausible, highly accurate segmentations of the carotid artery vessel wall. 
Moreover, the obtained segmentations are not only quantitatively accurate, but also plausible and clinically valuable. A qualitative analysis performed by an independent panel of six expert observers ranked our methods the highest in a composite score of four characteristics, outperforming all other challenge submissions in contour quality, bifurcation- and plaque segmentation, and segmentations in areas with flow artefacts. 

The use of 3D polar images or augmentation substantially improved the DSC for the vessel wall, and the median Hausdorff distances for the lumen. However, the median Hausdorff distance for the outer wall remained largely unchanged. A possible explanation is that the outer wall is much more challenging to detect accurately, due to the varying appearances of atherosclerotic plaque, containing calcified areas in some cases. Moreover, the contrast between the vessel wall and surrounding tissue is much lower than between the lumen and the vessel wall. We recommend using more data to cover the large variety of plaque appearances and improve overall segmentation performance.

Lastly, we have shown that our proposed method is effective for segmentation of vessel walls of carotid arteries. However, we observed that our method struggles with accurately segmenting the walls when there are horizontal displacements in the vessel, since ray-casting is performed in the axial plane. In this case, the Cartesian cross-section of the vessel is not circular. Since carotid arteries are mainly straight, the effect of this was limited. For using this algorithm for vessels with more curvature, we suggest using polar conversion on slices orthogonal to the centerline.

In conclusion, we have shown that it is feasible to automatically obtain anatomically plausible segmentations of the carotid vessel wall with high accuracy.

\section*{ACKNOWLEDGEMENTS}
Jelmer M. Wolterink was supported by NWO domain Applied and Engineering Sciences VENI grant (18192).
% The authors thank the organisation of the Carotid Artery Vessel Wall challenge\cite{Challenge} for making part of the CARE-II data set available for development and validation of our method.

%\textit{Results on the Carotid Artery Vessel Wall Segmentation Challenge will be presented orally at the SMRA 2021 conference and a MICCAI 2021 conference workshop. No written report on this work has or will be published elsewhere, and the systematic experimental comparison in this abstract will not be presented elsewhere.}

% References
\bibliography{report} % bibliography data in report.bib
\bibliographystyle{spiebib} % makes bibtex use spiebib.bst

\end{document}